\documentclass[MSNbibl,dvips]{arxstspdf}
\usepackage{flushend}
\usepackage{stfloats}

\volume{26}
\issue{2}
\pubyear{2011}
\firstpage{240}
\lastpage{256}
\doi{10.1214/10-STS346}

\begin{document}
\begin{frontmatter}

\title{Impact of Frequentist and Bayesian Methods on Survey
Sampling Practice: A~Selective Appraisal{\thanksref{T1}}}
\relateddoi{T1}{Discussed in \doi{10.1214/11-STS346A},
\doi{10.1214/11-STS346B} and \doi{10.1214/11-STS346C}; rejoinder at \doi{10.1214/11-STS346REJ}.}
\runtitle{Bayesian methods on Survey
Sampling}

\begin{aug}
\author{\fnms{J. N. K.} \snm{Rao}\corref{}\ead[label=e1]{jrao@math.carleton.ca}}

\runauthor{J. N. K. Rao}

\affiliation{Carleton University}

\address{J. N. K. Rao is Distinguished Research Professor, School of Mathematics and
Statistics, Carleton University, Ottawa, Ontario K1S 5B6, Canada
\printead{e1}.}

\end{aug}

\begin{abstract}
According to Hansen, Madow and Tepping [\textit{J. Amer. Statist. Assoc.} \textbf{78} (1983) 776--793],
``Probability sampling designs and randomization inference are widely
accepted as the standard approach in sample surveys.'' In this article,
reasons are advanced for the wide use of this design-based approach,
particularly by federal agencies and other survey organizations
conducting complex large scale surveys on topics related to public
policy. Impact of Bayesian methods in survey sampling is also discussed
in two different directions: nonparametric calibrated Bayesian
inferences from large samples and hierarchical Bayes methods for small
area estimation based on parametric models.
\end{abstract}


\end{frontmatter}

\section{Introduction}\label{1}

Sample surveys have long been conducted to obtain reliable estimates of
finite population descriptive parameters, such as totals, means, ratios
and quantiles, and associated standard errors and normal theory
intervals with large enough sample sizes. Probability-sampling designs
and randomization (repeated sampling) inference, also called the
design-based approach, played a dominant role, especially in the
production of official statistics, ever since the publication of the
landmark paper by Neyman (\citeyear{NEY34}) which laid the theoretical foundations
of the design-based approach. Neyman's approach was almost universally
accepted by practicing survey statisticians and it also inspired various
important theoretical contributions, mostly motivated by practical and
efficiency considerations. In this paper I~will first provide some
highlights of the design-based approach, for handling sampling errors,
to demonstrate its significant impact on survey sampling practice,
especially on the production of official statistics (Sections \ref{2} and
\ref{3.1}).

Model-dependent approaches (Section~\ref{3.2}) that lead to conditional
inferences more relevant and appealing than repeated sampling inferences
have also been advanced (Brewer, \citeyear{Bre63}; Royall, \citeyear{ROY70}). Unfortunately, for
large samples such approaches may perform very poorly under model
misspecifications; even small model deviations can cause serious
problems (\citeauthor{HANMADTEP83}, \citeyear{HANMADTEP83}). On the other hand, model-dependent
approaches can play a~vital role in small area (domain) estimation,
where the area-specific sample sizes are very small or even zero and
make the design-based area-specific direct estimation either very
unreliable or not feasible. Demand for reliable small area statistics
has greatly increased in recent years and to meet this growing demand,
federal statistical agencies and other survey organizations are
currently paying considerable attention to producing small area
statistics using models and methods that can ``borrow strength'' across
areas. Hierarchical Bayes (HB) model-dependent me\-thods are particularly
attractive in small area estimation because of their ability to handle
complex modeling and provide ``exact'' inferences on desired parameters
(Section \ref{5}). I will highlight some HB developments in small area
estimation that seem to have a significant impact on survey practice.
I~will also discuss the role of nonparametric Bayesian methods for
inferences, based on large area-specific sample sizes, especially those
providing Bayesian inferences that can be also justified under the
design-based framework (Section \ref{4.2}).

$\!\!$Models are needed, regardless of the approach used, to handle
nonsampling errors that include measurement errors, coverage errors and
missing data due to nonresponse. In the design-based approach, a
combined design and modeling approach is used to minimize the reliance
on models, in contrast to fully model-dependent approaches (Section
\ref{3.4}).

For simplicity, I will focus on descriptive parameters, but survey data
are also increasingly used for analytical purposes, in particular, to
study relationships and making inferences on model parameters under
assumed super-population models. For example, social and health
scientists are interested in fitting linear and logistic regression
models to survey data and then making inferences on the model parameters
taking account of the survey design features (Section \ref{3.3}).

\section{Design-Based Approach: Early Landmark Contributions}\label{2}

In this section I will highlight some early landmark contributions to
the design-based approach
that had major impact on survey practice. Prior to Neyman (\citeyear{NEY34}),
sampling was implemented either by ``balanced'' sampling through
purposive selection or by probability sampling with equal inclusion
probabilities. Such a method was called the ``representative method.''
Bowley (\citeyear{BOW26}) studied stratified \mbox{random} sampling with proportional
sample size allocation, leading to a representative sample with equal
\mbox{inclusion} probabilities. Neyman (\citeyear{NEY34}) broke through this restrictive
setup by relaxing the condition of equal \mbox{inclusion} probabilities and
introducing the ideas of efficiency and optimal sample size allocation
in his theory of stratified random sampling. He also demonstrated that
balanced purposive sampling may perform poorly if the underlying\vadjust{\eject} model
assumptions are violated. Neyman proposed normal theory confidence
intervals for large samples such that the frequency of errors in the
confidence statements based on all possible stratified random samples
that could be drawn does not exceed the limit prescribed in advance
``\textit{whatever the unknown properties of the finite population.}''
He broadened the definition of representative method by calling any
method of sampling that satisfies\vadjust{\goodbreak} the above frequency statement as
representative. It is interesting to note that Neyman advocated
distribution-free design-based inferences for survey sampling in
contrast to his own fundamental work on parametric inference, including
the Neyman--Pearson theory of hypothesis testing and confidence
intervals.

The possibility of developing efficient probability sampling designs by
minimizing total cost subject to a specified precision of an estimator
or maximizing precision for a given cost, taking account of operational
considerations, and making distribution-free inferences (point
estimation, variance estimation and large sample confidence intervals)
through the design-based approach were soon recognized.\break This, in turn,
led to a significant increase in the number and type of surveys taken by
probability sampling and covering large populations. In the early
stages, the primary focus was on sampling errors.

I now list a few important post-Neyman theoretical developments in the
design-based approach. As early as 1937, Mahalanobis used multistage
sampling designs for crop surveys in India. His classic 1944 paper
(Mahalanobis, \citeyear{MAH44}) presents a rigorous theoretical setup and a
generalized approach to the efficient design of sample surveys of
different crops in Bengal, India, with emphasis on variance and cost
functions. Mahalanobis considered a geographical region of finite area
and defined a field consisting of ``a finite number, say, $N_{0}$, of
basic cells arranged in a definite space or geographic order together
with a single value (or a set of values in the multivariate case) of
$z$ for each basic cell,'' where~$z$ is the variable of interest (say,
crop yield). Under this setup, he studied four different probability
sampling designs for selecting a sample of cells (called quads): unitary
unrestricted, unitary configurational, zonal unrestricted and zonal
configurational. In modern terminology, the four designs correspond to
simple random sampling, stratified random sampling, single-stage cluster
sampling and single-stage stratified cluster sampling, respectively. He
developed realistic cost functions depending on particular situations.
He also extended the theoretical setup to subsampling of\vadjust{\eject} clusters (which
he named as two-stage sampling). We refer the reader to Murthy (\citeyear{MUR64})
for a detailed account of the 1944 paper and other contributions of
Mahalanobis to sample surveys. Hall (\citeyear{Hal03}) provides a scholarly
historical account of the pioneering contributions of Mahalanobis to the
early development of survey sampling in India. Mahalanobis was
instrumental in establishing the National Sample Survey of India and the
famous Indian Statistical Institute.

Survey statisticians at the U.S. Census Bureau, under the leadership of
Morris Hansen, made fundamental contributions to survey sampling theory
and practice during the period 1940--1970, and many of those methods are
still widely used in practice. This period is regarded as the golden era
of the Census Bureau. Hansen and Hurwitz (\citeyear{HanHur43}) developed the basic
theory of stratified two-stage cluster sampling with one cluster or
primary sampling unit (PSU) within each stratum drawn with probability
proportional to a size measure (PPS) and then subsampled at a rate that
ensures self-weighting (equal overall probabilities of selection). This
method provides approximately equal interviewer work loads which are
desirable in terms of field operations. It can also lead to significant
variance reduction by controlling the variability arising from unequal
PSU sizes without actually stratifying by size and thus allowing
stratification on other variables to further reduce the variance. The
Hansen--Hurwitz method, with some modifications, has been widely used for
designing large-scale socio-economic, health and agricultural surveys
throughout the world. Many large-scale surveys are repeated over time,
such as the monthly Current Population Survey (CPS), and rotation
sampling with partial replacement of ultimate units (e.g., hou\-seholds)
is used to reduce response burden. Hansen et al. (\citeyear{HANHUR}) developed simple
but efficient composite estimators under rotation sampling in the
context of stratified multistage sampling. Rotation sampling and
composite estimation are widely used in large-scale surveys.

Prior to the 1950s, the primary focus was on estimating totals, means
and ratios for the whole popula\-tion and large planned subpopulations
such as US states or provinces in Canada. Woodruff (\citeyear{Woo52}) developed a
unified design-based approach for construc\-ting confidence intervals on
quantiles using only the estimated distribution function and the
associated standard error. This ingenious method is \mbox{applicable} to
general probability sampling designs and performs well in terms of
coverage probabilities in many cases.\vadjust{\eject} Woodruff intervals can also be
used to obtain standard errors of estimated quantiles (Rao and Wu, \citeyear{RaoWu87};
Francisco and Fuller, \citeyear{FraFul91}). Because of those features, the Woodruff
method had a signi\-ficant impact on survey practice. However, the~me\-thod
should not be treated as a black box for constructing confidence
intervals on quantiles, because it can perform poorly in some practical
situations. For example,~it performed very poorly under stratified
random sampling when the population is stratified by a conco\-mitant
variable $x$ highly correlated with the varia\-ble of interest $y$ (\citeauthor{KovRaoWu88},
\citeyear{KovRaoWu88}). The failure of the Woodruff method in this~ca\-se stems from
the fact that the standard error of the estima\-ted distribution function
at the quantile will be too small due to zero contributions to the
standard error from most strata. \citeauthor{KovRaoWu88} (\citeyear{KovRaoWu88}) showed that the
bootstrap method for stratified random sampling performs better than the
Woodruff method in this case, but in other situations the Wood\-ruff
method is better.

Attention was also given to inferences for unplan\-ned subpopulations
(also called domains) such as age--sex groups within a state. Hartley
(\citeyear{HAR59}) and Durbin (\citeyear{Dur58}) developed simple, unified theories for domain
estimation applicable to general designs, requiring only existing
formulae for population totals and means.

After the consolidation of basic design-based sampling theory, Hansen et
al. (\citeyear{HANetal51}) and others paid attention to measurement errors in surveys.
They developed basic theories under additive measurement error models
with minimal model assumptions on the observed responses treated as
random variables. Total variance of an estimator is decomposed into
sampling variance, simple response variance and correlated response
variance (CRV) due to interviewers. The CRV was shown to dominate the
total variance when the number of interviewers is small, and the 1950
U.S. Census interviewer variance validation study showed that this
component is indeed large for small areas. Partly for this reason,
self-enumeration by mail was first introduced in the 1960 U.S. Census to
reduce the CRV component. Earlier, Mahalanobis (\citeyear{MAH46}) developed the
method of interpenetrating subsamples (called replicated sampling by
Deming, \citeyear{Dem60}) and used it extensively in large-scale surveys in India
for assessing both sampling and interviewer errors. By assigning the
subsamples at random to interviewers, the total variance can be
estimated and interviewer differences assessed.\vadjust{\eject}

It should be clear from the above brief description of early
developments that much of the basic sampling theory was developed by
official statisticians or those closely associated with official
statistics. Theory was driven by the need to solve real problems and
often theory was not challenging enough to attract academic researchers
to survey sampling. As a~result, university researchers paid little
attention to survey sampling in those days with few exceptions (e.g.,
Iowa State University under the leadership of Cochran, Jessen and
Hartley).

\section{Some Recent Design-Based and Non-Bayesian
Developments}\label{3}

\subsection{Model-Assisted Approach}\label{3.1}

We first give a brief account of the model-assisted approach that uses a
working model to find efficient estimators. However, the associated
inferences are design-based. Consider a finite population $U$ consisting
of $N$ elements labeled $1,\ldots,N$ with associated values
$y_{1},\ldots,y_{N}$ of a variable of interest $y$. Under a probability
sampling design, the inclusion probabilities $\pi_{1},\ldots,\pi_{N}$ are
all strictly positive and a basic estimator of the total $Y = \sum_{i
\in U} y_{i}$ is of the form $\hat{Y} = \sum_{i \in s} d_{i} y_{i}$, where
$s$ denotes a sample and $d_{i} = \pi_{i}^{ - 1}$ are the so-called
design weights (Horvitz and Thompson, \citeyear{HorTho52}; Narain, \citeyear{Nar51}). For example,
in the Neyman stratified random sampling design, the design weights are
equal to the inverse of the sampling fractions within strata and vary
across strata, while in the Hansen et al. two-stage cluster sampling
design, the design weights are all equal. Design unbiasedness of
estimators is not insisted upon (contrary to statements in some papers
on inferential issues of sampling theory) because it ``often results in
much larger MSE than necessary'' (\citeauthor{HANMADTEP83}, \citeyear{HANMADTEP83}). Instead, design
consistency is deemed necessary for large samples. Strategies (design
and estimation) that appea\-red reasonable are entertained (accounting for
costs) and relative properties are carefully studied by analytical and/or empirical methods, mainly through the comparison of mean squared
error (MSE) or anticipated MSE under plausible population models on the
variables $y_{i}$ treated as random variables. This is essentially the
basis of the repeated sampling (or designbased) approach.

In recent years, a model-assisted repeated sampling approach has
received the attention of survey practitioners. In this approach, a
working population model is used to find efficient design-consistent
estimators. For example, suppose the working model is a linear
regression model of the form
\begin{equation}\label{eq1}
y_{i} = x'_{i}\beta + \varepsilon_{i};\quad i = 1,\ldots,N,
\end{equation}
with model errors $\varepsilon_{i}$ assumed to be uncorrelated with mean
zero and variance proportional to a known constant $q_{i}$, where
$x_{i}$ is a vector of auxiliary variables with known population
total $X$. Under mo\-del~(\ref{eq1}), the best linear unbiased estimator (BLUE) of
the model parameter $\beta$, based on the census values $\{
(y_{i},x_{i});i \in U\}$, is given by the ``census'' regression
coefficient
\[
B = \biggl(\sum_{i \in U} x_{i} x'_{i}/q_{i}\biggr)^{ - 1}\biggl(\sum_{i \in U} x_{i}
y_{i}/q_{i}\biggr).
\]
A predictor of $y_{i}$ under the working model is then given by
$\hat{y}_{i} = x'_{i}\hat{B}$ for $i = 1,\ldots,N$ where $\hat{B}$ is the
design-weighted estimator of $B$:
\[
\hat{B} = \biggl(\sum_{i \in s} d_{i} x_{i}x'_{i}/q_{i}\biggr)^{ - 1}\biggl(\sum_{i \in
s} d_{i} x_{i}y_{i}/q_{i}\biggr).
\]
By writing the total as $Y = \sum_{i \in U} \hat{y}_{i} + \sum_{i \in U}
e_{i}$ where $e_{i} = y_{i} - \hat{y}_{i}$ denotes the prediction error, a
design-based estimator of $Y$ is given by $\sum_{i \in U} \hat{y}_{i} +
\sum_{i \in s} d_{i} e_{i}$. We can express this estimator as a
generalized regression (GREG) estimator
\begin{equation}\label{e2}
\hat{Y}_{\mathrm{gr}} = \hat{Y} + \hat{B}'(X - \hat{X}),
\end{equation}
where $\hat{X} = \sum_{i \in s} d_{i} x_{i}$ (S{\"a}rndal, Swensson and Wretman, \citeyear{SarSweWre92}).
The
GREG estimator (\ref{e2}) is design-consis\-tent regardless of the validity of
the working model (Robinson and Sarndal, \citeyear{RobSar83}) under certain regularity
conditions provided $X$ is precisely correct. If the working model
provides a good fit to the data, then the residuals $e_{i}$ should be
less variable than the response values $y_{i}$ and the GREG estimator is
likely to be significantly more efficient than the basic design-weighted
estimator~$\hat{Y}$.

The estimator (\ref{e2}) may also be expressed as a weigh\-ted sum $\sum_{i \in s}
w_{i} y_{i}$, where $w_{i} = d_{i}g_{i}$ with
\begin{equation}\label{e3}
g_{i} = 1 + (X - \hat{X})'\biggl(\sum_{i \in s} d_{i} x_{i}x'_{i}/q_{i}\biggr)^{ -
1}x_{i}q_{i}^{ - 1}.
\end{equation}
The adjustment factors $g_{i}$, popularly known as the~$g$-weights,
ensure the calibration property $\sum_{i \in s} w_{i} x_{i} = X$ so that
the GREG estimator when applied to the sample values $x_{i}$ agrees with
the known total~$X$. This property is attractive to the user when the
vector $X$ contains user-specified totals.

The assumption of a working linear regression mo\-del (\ref{eq1}) can be relaxed
by adopting more flexible working models. For example,
Breidt, Claeskens and Opsomer
(\citeyear{BreClaOps05}) proposed a nonparametric model-assisted approach based on a
penalized spline (P-spline) regression working model and showed that the
resulting estimators are design-consistent and more efficient than the
usual GREG estimators ba\-sed on linear regression working models when the
latter are incorrectly specified. Also, the P-spline model-assisted
estimators were shown to be approximately as efficient as the GREG
estimators when the linear regression working model is correctly
specified. The P-spline approach can be easily implemented using
existing estimation packages for GREG because the underlying model is
closely related to a~linear regression model. It offers a wider scope
for the model-assisted approach because it makes minimal assumptions on
the regression of $y$ on $x$ without assuming a specific parametric form.

Under the model-assisted approach, design-consis\-tent variance estimators
are obtained either by a~Tay\-lor linearization method or by a resampling
me\-thod (when applicable), provided the probability sampling design
ensures strictly positive joint inclusion probabilities $\pi_{ij}$, $i \ne
j$. Using the estimator and associated standard error, asymptotically
valid normal theory intervals are obtained regardless of the validity of
the working model.

Most large-scale surveys are multipurpose and observe multiple variables
of interest, and the same working model may not hold for all the
variables of interest. In that case, a model-assisted approach may lead
to possibly different $g_{i}$ and hence different calibration
weights $w_{i}$ associated with the variables, that is, the calibration
weights are of the form~$w_{ij}$ associated with the variable $j$ and
sample unit~$i$. However, survey users prefer to use a common\break weight~%
$w_{i}$ for all variables of interest. This is often accomplished by
minimizing a suitable distance measure between $d_{i}$ and $w_{i}$ for $i
\in s$ subject to userspecified calibration constraints, say, $\sum_{i \in
s} w_{i} z_{i} = Z$, without appealing to any working model, where
$Z$ is the vector of known totals associated with the user-specified
variables $z$. For example, a chi-squared distance measure leads to
common calibration\break weights~$w_{i}$ of the form $d_{i}g_{i}$ where $g_{i}$ is given by~(\ref{e3})
with~$x_{i}$ replaced by $z_{i}$ (Deville and Sarndal, \citeyear{DevSar92}). Thus, calibration
estimation in this case corresponds to using model-assisted estimation
based on a linear regression model (\ref{eq1}) with $z_{i}$ as the vector of
predictor variables. Calibration estimation has attracted the attention
of users due to its ability to produce common calibration weights and
accommodate an arbitrary number of user-specified calibration (or
benchmark) constraints, for example, calibration to the marginal counts
of several post-stratification va\-riables. Several national statistical
agencies have developed software designed to compute calibration
weights: GES (Statistics Canada), LIN WEIGHT (Statistics Netherlands),
CALMAR (INSEE, France) and CLAN97 (Statistics Sweden). Sarndal (\citeyear{SAR07})
says, ``Calibration has established itself as an important
methodological instrument in large-scale production of statistics.''
Brakel and Bethlehem (\citeyear{vanBET08}) noted that the use of common calibration
weights for estimation in multipurpose surveys makes the calibration
method ``very attractive to produce timely official releases in a
regular production environment.''

$\!\!$Unfortunately, the model-free calibration approach can lead to erroneous
inferences for some of the response variables, even in fairly large
samples if the underlying working linear regression model uses an
incorrect or incomplete set of auxiliary variables, unlike the
model-assisted approach that uses a working model obtained after some
model checking. For example, suppose that the underlying model is
a~qua\-dratic regression model of $y$ on $x$ and the distribution of $x$ is
highly skewed. Also, suppose that the user-specified calibration
constraints are the known population size $N$ and the known population
total $X$. In this case, the calibration estimator of the total $Y$ under
simple random sampling is the familiar simple linear regression
estimator with $x$ as the predictor variable. On the other hand, a
model-assisted estimator under the quadratic regression working model is
given by a multiple linear regression estimator with $x_{1} = x$ and
$x_{2} = x^{2}$ as the predictor variables, assuming the total of
$x_{2}$ is also known. \citeauthor{RAOJOCHID03} (\citeyear{RAOJOCHID03}) demonstrated that the coverage
performance of the normal theory interval associated with the
calibration estimator is poor even in fairly large samples, unlike the coverage
performance of the normal theory intervals associated with the
model-assisted estimator. The coverage performance depends on the
skewness of the residuals from the fitted model and in the case of
calibration estimation the skewness of residuals after fitting simple
linear regression remains large, whereas the skewness of residuals after
fitting quadratic regression is small even if $y$ and $x$ are highly
skewed. This simple example demonstrates that the population structure
does matter in design-based inferences and that it should be taken into
account through a model-assisted approach based on suitable working
models. But the model-assisted approach has the practical limitation
that the weight~$w_{i}$ may vary across variables in surveys with
multiple variables of interest, unlike in the calibration approach.
Also, for complex working models, such as the P-spline, all the
population values of the predictor variables should be known in order to
implement the model-assisted approach.

The model-assisted approach is essentially design-based, unlike the
model-dependent approach (Section~\ref{3.2}) that can provide conditional
inferences referring to the particular sample, $s$, of units selected.
Such conditional inferences may be more relevant and appealing than the
unconditional repeated sampling inferences used in the design-based
approach.

\subsection{Model-Dependent Approach}\label{3.2}

The frequentist model-dependent approach to inference assumes that the
population structure obeys a specified population model and that the
same mo\-del holds for the sample, that is, no sample selection bias with
respect to the assumed population model. Sampling design features are
often incorporated into the model to reduce or eliminate the sample
selection bias (see Section \ref{3.3} for some difficulties to implement this
in practice). Typically, distributional assumptions are avoided by
focusing on point estimation, variance estimation and associated normal
theory confidence intervals, as in the case of design-based inferences.
As a result, models used specify only the mean function and the variance
function of the variable of interest, $y$.We refer the reader to
Valliant, Dorfman and Royall (\citeyear{ValDorRoy00}) for an excellent account of the model-dependent approach.

As noted in Section \ref{1}, model-dependent strategies may perform poorly in
large samples when the population model is not correctly specified; even
small deviations from the assumed model that are not easily detectable
through routine model checking can cause serious problems. In the Hansen, Madow and Tepping (\citeyear{HANMADTEP83}) example of an incorrectly specified population model, the
best linear unbiased prediction (BLUP) estimator of the mean is not
design consistent under their stratified simple random sampling design
with near optimal sample allocation (commonly used to handle highly
skewed populations such as business populations). As a result,
model-dependent confidence intervals exhibited poor performance: for $n =
100$, coverage was around 70\% compared to nominal level of 95\%, while
coverage for model-assisted intervals was 94.4\%. To get around this
difficulty, Little (\citeyear{LIT83}) proposed restricting attention to models that
hold for the sample and for which the BLUP estimator is design
consistent. For example, in the \citeauthor{HANMADTEP83} (\citeyear{HANMADTEP83}) example, the BLUP of the
mean under a model with means differing across strata is identical to
the traditional stratified mean which is design consistent. But it seems
not possible even to find a suitable model under which the widely used
combined ratio estimator of the mean under the stratified random
sampling is the BLUP estimator. The combined ratio estimator is a
model-assisted estimator under a~ratio working model with a common
slope. It allows a~large number of strata with few sample units from
each stratum and yet remains design consistent, unlike the separate
ratio estimator which is the BLUP estimator under a ratio model with
separate slopes across strata: $E(y_{hi}|x_{hi}) =
\beta_{h}x_{hi}$, $V(y_{hi}|x_{hi}) = \sigma^{2}x_{hi}$, where $y_{hi}$ and
$x_{hi}$ denote the values of the variable of interest $y$ and an
auxiliary variable $x$ for the unit $i$ in stratum $h$. Moreover, the BLUP
estimator under this model requires the knowledge of the strata
population means
$\bar{X}_{h}$, whereas the combined ratio estimator requires only the
overall population mean~$\bar{X}$.

It is also not clear how one proceeds to formulate suitable models for
general sampling designs that lead to design-consistent BLUP estimators.
Further, the main focus has been on point estimation and it is not clear
how one should proceed with variance estimation and setting up
confidence intervals that have repeated sampling validity. In this
context, Pfeffermann (\citeyear{PFE09}) says, ``I presume that these are supposed to
be computed under the corrected model as well. Are we guaranteed that
they are sufficiently accurate under the model? Do we need to robustify
them separately?'' Little (\citeyear{Lit08}), in his rejoinder to Pfeffermann's
comment, says that he advocates using some replication method for
variance estimation and then appealing to normal approximation for
confidence intervals. Clearly, further work is needed to address the
above issues. Note that if parametric assumptions are made, such as
normality of model errors, then it is possible to make exact Bayesian
inferences by introducing suitable priors on the model parameters
(Section \ref{4}).

Some recent work on the model-dependent approach focused on avoiding
misspecification of the mean function $E(y|x) = m(x)$ by using
P-spline\vadjust{\eject}
models. Zheng and Little (\citeyear{ZHELIT03}, \citeyear{ZHELIT05}) studied single stage PPS sampling
using a P-spline model, based~on the size measure $x$ used in PPS
sampling, to repre\-sent the regression function $m(x)$, and a specified
function of the size measure as the variance function. In a simulation
study, they compared the performance of the usual linear GREG estimator
and the P-spline model-based estimators, not necessari\-ly
design-consistent, and showed that the P-spline~mo\-del-based estimators
are generally more efficient than the GREG or the NHT estimator in terms
of design MSE even for large samples. However, the simula\-tion study did
not consider model-assisted estimators corresponding to their P-spline
model. The simu\-lations also showed that the design-bias for their~%
P-spline estimators is minor, even though the estimators are not
design-consistent, and hence the authors conclude that ``design
consistency may not be of paramount importance.'' On the other hand, in
the \citeauthor{BreClaOps05} (\citeyear{BreClaOps05}) \mbox{simulation}\break study,~their model-assisted P-spline
estimator is so\-metimes much better, and never worse, than the
corresponding P-spline model-based estimator under stratified random
sampling. The latter estimator is not design-consistent under the
P-spline model considered by \citeauthor{BreClaOps05} (\citeyear{BreClaOps05}).

As noted above, a main advantage of the frequentist model-dependent
approach is that it leads to inferences conditional on the selected
sample of units, $s$, unlike the unconditional design-based approach.
However, it is possible to develop a conditional mo\-del-assisted approach
that allows us to restrict the reference set of samples to a
``relevant'' subset of all possible samples specified by the design.
Conditionally valid inferences for large samples can then be obtained.
Rao (\citeyear{RAO92}) and Casady and Valliant (\citeyear{CASVAL93}) developed an ``optimal''
linear regression estimator that is asymptotically valid under the
conditional setup.

We refer the reader to Kalton (\citeyear{KAL02}) for compe\-lling arguments for
favoring design-based approaches (possibly model-assisted and/or
conditional) to handle sampling errors. Smith (\citeyear{SMI94}) named the
traditional repeated sampling inference as ``procedural inference'' and
argued that procedural inference is the correct approach for surveys in
the public domain.

\subsection{Analysis of Complex Survey Data}\label{3.3}

Data collected from large-scale socio-economic,\break health and other surveys
are being extensively used for analysis purposes, such as inferences on
the\vadjust{\eject} regression parameters of linear and logistic regression population
models. Ignoring the survey design features and using standard methods
can lead to erroneous \mbox{inferences} on model parameters because of sample
selection bias caused by informative sampling. It is tempting to expand
the models by including among the predictors all the design variables
that define the selection process at the various levels and then ignore
the design and apply standard methods to the expanded model. The main
difficulties with this approach, advocated by some leading researchers,
are the following, among others (Pfeffermann and Sverchkov, \citeyear{PfeSve03}): (1)~Not all design variables may be known or accessible to the analyst. (2)
Too many design variables can lead to difficulties in making inferences
from the expanded models. (3)~The expanded model may no longer be of
scientific interest to the analyst.

The design-based approach can provide asymptotically valid repeated
sampling inferences without changing the analyst model. A unified
approach based on survey-weighted estimating equations leads to
design-consistent estimators of the ``census'' or finite population
parameters, which in turn estimate the associated model parameters.
Further, using resampling methods for variance estimation, such as the
jackknife and the bootstrap for survey data, asymptotically valid
design-based inferences on the census parameters can be implemented. The
same methods may also be applicable for inference on the model
parameters, in many cases of large-scale surveys. In the other cases, it
is necessary to estimate the model variance of the census parameters
from the sample. The estimate of the total variance is then given by the
sum of this estimate and the resampling variance estimate.

$\!\!$In practice, the data file would contain for each~sam\-pled unit the
variables of interest and predictor variables, final weights after
adjustment for unit nonresponse and the corresponding replication
weights, for example, bootstrap weights. The analyst can use software
that handles survey weights (such as SAS) to obtain point estimates from
the final weights and the corresponding point estimates for each
bootstrap replicate using bootstrap weights. The variabi\-lity of the
bootstrap point estimates provides asymptotically valid standard errors
for designs commonly used in large-scale surveys. Details of the methods
are not provided due to space limitations, but the reader is referred to
Rao (\citeyear{RAO05}, Section 6) for a succinct account of analysis of survey data
using resampling methods for variance estimation and\vadjust{\eject} normal theory
confidence intervals. Design-based approach using resampling methods is
extensively used in practice and software is also available (e.g.,
WesVar, Stata).

The design-based approach has also been applied to make inferences on
the regression parameters and the variance parameters of multilevel
models from data obtained from multistage sampling designs corresponding
to the levels of the model. For example, in an education study of
students, schools (first-sta\-ge units or clusters) may be selected with
probabilities proportional to school size and students (second-stage
units) within selected schools by stratified random sampling. Again,
ignoring the sampling design and using traditional methods for
multilevel models that ignore the design can lead to erroneous
inferences in the presence of sample selection bias. In~the design-based
approach, estimation of variance parameters of the model is more
difficult than that~of regression parameters and the necessary
information for estimating variance parameters is often not provided in
public-use data files which typically report only the final weight for
each sample unit. Widely used design-based methods have been proposed in
the literature (e.g., Pfeffermann et al., \citeyear{Pfeetal98}, and Rabe-Hesketh and
Skrondal, \citeyear{RabSkr06}) to handle variance parameters that require the weights
within sampled clusters in addition to the weights associated with the
clusters. Some of those methods can be implemented using the Stata
program gllamm. Unfortunately, the resulting estimators of variance
parameters may not be design-model consistent when the sample sizes
within clusters are small, even for two-level linear models. Korn and
Graubard (\citeyear{KorGra03}) demonstrated the bias problem and proposed a different
method for simple two-level or three-level models involving only a
common mean as the fixed effect. This method first obtains the census
parameters and then estimates those parameters. It worked well in
empirical studies even for small within-cluster sample sizes. Rao, Verret and Hidiroglou
(\citeyear{RAOVERHID}) proposed a weighted estimating equations (WEE) approach for
general two-level linear models that uses within-cluster joint inclusion
probabilities, similar to Korn and Graubard (\citeyear{KorGra03}). The WEE method leads
to design-model consistent estimators of variance parameters even for
small\break within-cluster sample sizes, provided the number of sample
clusters is large. It performed well in empirical studies compared to
the other methods proposed in the literature. \citeauthor{RAOVERHID}
(\citeyear{RAOVERHID}) also
proposed a unified approach based on a weighted log-composite likelihood
that can handle generalized linear multilevel models and small
within-cluster sample sizes. This method is currently under
investigation.

A drawback of the design-based approach to the analysis of survey data
is that it may lead to loss in efficiency when the final weights vary
considerably across the sampling units. Alternative approaches that can
reduce the variability of the weights and thus lead to more efficient
estimators have also been proposed (e.g., Pfeffermann and Sverchkov,
\citeyear{PfeSve03}; Fuller, \citeyear{FUL09}, Chapter 6). We refer the reader to Pfeffermann
(\citeyear{PFE93}) and Rao et al. (\citeyear{RAOetal10}) for overviews on the role of sampling
weights in the analysis of survey data.

\subsection{Nonsampling Errors}\label{3.4}

Survey practitioners have to rely on models, regardless of the approach
used, to handle nonsampling errors that include measurement errors,
coverage errors and missing data due to unit nonresponse and item
nonresponse. In the design-based approach, a combined design and
modeling approach is used to minimize the reliance on models, in
contrast to fully model-dependent approaches that will have similar
difficulties noted in the previous subsections. As mentioned in Section
\ref{1}, Hansen et al. (\citeyear{HANetal51}) studied measurement errors under minimal model
assumptions on the observed responses trea\-ted as random variables, and
their discovery that the correlated response variance due to
interviewers dominates the total variance when the number of
interviewers is small led to the adoption of self enumeration by mail in
the 1960 U.S. Census.

Inference in the presence of missing survey data, particularly item
nonresponse, has attracted a lot of \mbox{attention}; see Little and Rubin
(\citeyear{LitRub02}) for an excellent account of missing data methods. To handle item
nonresponse, imputation of missing data is often used because of its
practical advantages. In the design-based approach, traditional weighted
estimators of a total or a mean are computed from the completed data
set, leading to an imputed estimator. Often imputed values are generated
from an imputation model assumed to hold for the respondents under a
missing at random (MAR) response mechanism. Under this setup, the
imputed estimator is unbiased or asymptotically unbiased under the
combined design and model set up. Reiter,
Raghunathan and Kinney (\citeyear{REIRAGKIN06}) demonstrated the
importance of incorporating sampling design features into the imputation
model in order to make the model hold for the sample and then for the
respondents under the assumed MAR response mechanism. An alternative
approach avoids imputation models but assumes a model for the response
mechanism. For example, a popular method consists of forming imputation
classes (according to the values of estimated response probabilities
under a specified response model) and assuming that the missing values
are missing completely at random (MCAR) within classes. The missing
values are then imputed by selecting donors at random from the observed
values within classes. It may be also possible to develop imputation
methods that make an imputed estimator doubly robust in the sense that
it is valid either under an assumed imputation model or under an assumed
response mechanism (e.g., see Haziza and Rao, \citeyear{HAZRAO06}). Doubly robust
estimation has attracted considerable attention in the nonsurvey
literature (see, e.g., \citeauthor{CAOTSIDAV09}, \citeyear{CAOTSIDAV09}).

Variance estimation under imputation for missing survey data has
attracted a lot of attention because treating the imputed values as if
observed and then applying standard variance formulae can often lead to
serious underestimation because the additional variability due to
estimating the missing values is not taken into account. Methods that
can lead to asymptotically valid variance estimators under single
imputation for missing data have been proposed under the above setups.
We refer the reader to Kim and Rao (\citeyear{KIMRAO09}) for a unified approach to
variance estimation under the imputation model approach, and to Haziza
(\citeyear{Haz09}) for an excellent overview of imputation for survey data and
associated methods for inference. Rubin (\citeyear{Rub87}) proposed multiple
imputation to account for the underestimation when applying standard
formulae treating the imputed values as if observed. Under this
approach, $M$ ($\ge 2$) imputed values are generated for a missing item,
leading to $M$ completed data sets. Rubin recommends the use of
traditional design-based estimators and variance estimators, computed
from each of the completed data sets, although multiple imputation ideas
are based on a Bayesian perspective: ``We restrict attention to standard
scientific surveys and standard complete data statistics'' (Rubin, \citeyear{Rub87},
page~113). Multiple imputation estimator~$\hat{Y}_{\mathrm{MI}}$ of a total $Y$ is
taken as the average of the~$M$ estimators $\hat{Y}_{I1},\ldots,\hat{Y}_{IM}$, and its estimator of
variance is given by $v_{\mathrm{MI}} = \bar{v}_{M} + (1 + M^{ - 1})b_{M}$,
where~$\bar{v}_{M}$ is the average of the $M$ na\"{\i}ve variance estimators
$v_{I1},\ldots,v_{IM}$ and $b_{M} \!=\! (M - 1)^{ - 1}\sum_{m = 1}^{M}
(\hat{Y}_{Im} - \hat{Y}_{\mathrm{MI}})^{2}$. Rubin
gives design-based conditions for ``proper imputation'' that ensure the
repeated sampling validity of the estimator $\hat{Y}_{\mathrm{MI}}$ and the
associated variance estimator $v_{\mathrm{MI}}$ under a posited response
mechanism. Unfortunately, there are some difficulties in developing
imputation methods satisfying Rubin's conditions for ``proper
imputation'' with complex survey data (see, e.g., Kim et al., \citeyear{Kimetal06}).
Nevertheless, multiple imputation shows how Bayesian ideas can be
integrated to some extent with the traditional~de\-sign-based approach
that is widely used in practice.

\section{Bayesian Approaches}\label{4}

This section provides an account of both parametric and nonparametric
Bayesian (and pseudo-Bayesian) approaches to inference from survey data,
focusing on descriptive finite population parameters.

\subsection{Parametric Bayesian Approach}\label{4.1}

As noted in Section \ref{3.2}, the frequentist model-dependent approach mostly
avoided distributional assumptions by specifying only the mean function
and the variance function of the variable of interest. Under a specified
distribution on the assumed model, Bayesian inferences can be easily
implemen\-ted, provided the model holds for the sample. Royall and
Pfef\-fermann (\citeyear{RoyPfe82}) studied Bayesian inference on the population mean
assuming normality and flat (diffuse) priors on the parameters of a
linear regression model. Their focus was on the posterior mean and the posterior
variance and, hence, results were similar to those of Royall (\citeyear{ROY70})
without the normality assumption and priors on model parameters.
However, exact credible intervals on the mean and other parameters of
interest can be obtained conditional on the observed data, using a~%
pa\-rametric Bayesian setup. It can be implemented even under complex
modeling, using powerful Monte Carlo Markov chain (MCMC) methods to
simulate samples from the posterior distributions of interest. Scott and
Smith (\citeyear{SCOSMI69}) obtained the posterior mean and the posterior variance of
the population mean under linear models with random effects, normality
and diffuse priors on the model parameters. Their posterior mean is also
the BLUP estimator without the normality assumption when the variance
parameters of the model are known. In the frequentist approach,
estimates of variance parameters are substituted in the BLUP to get the
empirical BLUP (EBLUP) estimator which is different (but close to) to
the posterior mean. However, the Bayesian approach also provides the
posterior variance which is typically different from the estimated mean
squared prediction error (MSPE) of the EBLUP estimator; several
different methods of estimating MSPE have been proposed in the context
of small area estimation (Rao, \citeyear{Rao03}, Chapter 7). A simulation study by
Bellhouse and Rao (\citeyear{BelRao86}) showed that any gain in efficiency of the
posterior mean (or the BLUP) over traditional design-based estimators is
likely to be small in practice. However, by regarding the clusters as
small areas of interest, the Scott--Smith approach provides models
linking the small areas and the resulting estimators of small area means
can lead to significant efficiency gains over direct area-specific
estimators. Random cluster effect models are now extensively used to
construct efficient small area estimators by ``borrowing strength''
across small areas using auxiliary information (Section \ref{5}). It may be
noted that the empirical Bayes (EB) approach to inference from random
cluster effect models is similar to EBLUP, but it can handle general
parametric random cluster effect models and does not require the
linearity assumption used in the BLUP method. The EB approach is
essentially frequentist, unlike the Bayesian approach that requires the
specification of priors on the model parameters. It may be more
appropriate to name ``empirical Bayes'' as ``empirical best'' without
changing the abbreviation EB (Jiang and Lahiri, \citeyear{JiaLah06N1}).

$\!\!$Sedransk (\citeyear{SED77}) studied regression models with~random
slopes $\beta_{1},\ldots,\beta_{L}\sim_{\mathrm{i.i.d.}}N(\nu,\sigma_{\beta} ^{2})$,
using a prior distribution on $\nu$ specified
as $N(\beta_{0},\sigma_{0}^{2})$. He then followed the Scott and Smith
(\citeyear{SCOSMI69})\vadjust{\goodbreak} approach and obtained posterior mean and posterior variance of
the finite population total. He applied the method to data on banks from
the U.S. Federal Reserve Board to estimate a current monetary total
making use of extensive historical data to specify the~va\-lues of
$\beta_{0}$, $\sigma_{0}^{2}$ and thus arrive at an informative prior which
in turn leads to more efficient posterior inferences compared to those
based on a nonin\-formative prior, provided the informative prior is
correctly specified. Malec and Sedransk (\citeyear{MalSed85}) extended Scott--Smith
results to three-stage sampling. \citeauthor{NANSEDSMI97} (\citeyear{NANSEDSMI97}) applied the
Bayesian approach to obtain order-restricted estimators of the age
composition of a population of Atlantic cod, using MCMC methods.
Sedransk (\citeyear{SED08}) lists possible uses of parametric Bayesian methods for
sample surveys, including the above application\vadjust{\eject} to estimation from
establishment surveys, ``optimal'' sample allocation and small area
estimation from~da\-ta pooled from independent surveys (see Section~\ref{5}).

\citeauthor{PfeMouSil06} (\citeyear{PfeMouSil06}) report an interesting application of the
Bayesian approach to make inferences from multilevel models under
informative sampling. In this case, the multilevel sample model induced
by informative sampling is more complicated than the corresponding
population model and, as a result, frequentist methods are difficult to
implement. On the other hand, the authors show that the Bayesian
approach, using noninformative priors on the model parameters indexing
the sample model and applying MCMC methods, is efficient and convenient
for handling such complex sample models, although computer intensive.
This application is an example where the Bayesian approach offers
computational advantage over the corresponding frequentist approach.

\subsection{Nonparametric Bayesian Approaches}\label{4.2}

For multipurpose large-scale surveys, parametric Bayesian methods based
on distributional assumptions have limited value because of the
difficulties in validating the parametric assumptions. It may be more
appealing to use a nonparametric Bayesian approach, but this requires
the specification of a~nonparametric likelihood function based on the
full sample data $\{ (i,y_{i}),i \in s\}$ and a prior distribution on the
parametric vector $(y_{1},\ldots,y_{N})$. The likelihood function based on
the full sample data, however, is noninformative in the sense that all
possible unobserved values of the parameter vector have the same
likelihood function (Godambe, \citeyear{God66}). One way out of this difficulty is
to take a Bayesian route by assuming an informative (exchangeable) prior
on the $N $-dimensional parameter vector and combine it with the
noninformative likelihood (Ericson, \citeyear{Eri69}; Binder, \citeyear{Bin82}) to get an
informative posterior, but inferences do not depend on the sample
design; Ericson argued that an exchangeable prior assumption may be
reasonable under simple random sampling. Ericson (\citeyear{Eri69}) focused on the
posterior mean and the posterior variance of the population
mean $\bar{Y}$ which approximately agree, under prior vagueness, with the
usual formulae under the design-based approach. In the case of
stratified sampling with known strata differences, priors within strata
are assumed to be exchangeable.

Meeden and Vardeman (\citeyear{MeeVar91}) used a Polya posterior (PP) over the
unobserved, assuming that ``unseen are like the seen'' (equivalent to\vadjust{\eject}
exchangeability). In this case, the posterior ``does not arise from a~%
single prior distribution'' (Meeden, \citeyear{MEE95}) and, hen\-ce, it is called a
pseudo-posterior. It is also similar to the Bayesian bootstrap (Rubin,
\citeyear{Rub81}; Lo, \citeyear{Lo88}). The Polya posterior is a flexible tool and methods
based on PP have reasonable design-based properties under simple random
sampling. PP approach permits Bayesian interval estimation for the mean
and any other parameters of interest through simulation of many finite
populations from PP. The general interval estimation feature of the PP
approach is attractive. Meeden (\citeyear{MEE95}) extended the PP approach to
utilize auxiliary population information $(x_{1},\ldots,x_{N})$ by making a
strong prior assumption that the ratios $r_{i} = y_{i}/x_{i}$ are
exchangeable and obtained point and interval estimators for the
population median. Empirical results under simple random sampling are
given to show that the resulting Bayesian intervals perform well in
terms of design-based coverage. \citeauthor{LAZMEENEL08} (\citeyear{LAZMEENEL08}) developed a
constrained Polya posterior to generate simulated populations that are
consistent with the known population mean of an auxiliary variable $ x$,
using MCMC methods. This approach permits the use of known population
auxiliary information and leads to more efficient Bayesian inferences.
Nelson and Meeden (\citeyear{NelMee98}) adapted PP to incorporate prior knowledge that
the population median belongs to some interval. Meeden (\citeyear{Mee99}) studied
two-stage cluster sampling (balanced case) and his two-stage PP-based
results for the posterior mean and the posterior variance are very close
to standard design-based results, but it is not clear how readily
Meeden's approach extends to the unbalanced case with unequal cluster
sizes. Although the PP approach is attractive and seems to provide
calibrated Bayesian inferences at least for some simple sampling
designs, it is unlikely to be used in the production of official
statistics because of the underlying assumption that ``unseen are like
the seen'' and each case needs to be studied carefully to develop
suitable PP. Also, it is not clear how this method can handle complex
designs, such as stratified multistage sampling designs, or even single
stage unequal probability sampling without replacement with
nonnegligible sampling fractions, and provide design-calibrated
Bayesian inferences. \mbox{Nevertheless}, the PP approach may be useful for
some specialized surveys and when inferences are desired on a variety of
finite population parameters associated with the variable of interest
$y$ or prior knowledge on the parameters is available, as in the case of
Nelson and Meeden (\citeyear{NelMee98}).

An alternative approach is to start with an informative likelihood based
on reduced data. For example, under simple random sampling, it may be
reasonable to suppress the labels $i$ from the full data $\{ (i,y_{i}),i
\in s\}$ and use the likelihood based on the reduced data $\{ y_{i},i \in
s\};$ see Hartley and Rao (\citeyear{HARRao68}) and Royall (\citeyear{ROY68}). On the other hand,
for stratified random sampling, labels within strata are suppressed but
strata labels are retained because of known strata differences. Hartley
and Rao (\citeyear{HARRao68}) proposed a ``scale-load'' approach for inference on the
mean $\bar{Y}$. Under this approach, the $y $-values are assumed to
belong to a finite set of possible values $\{ h_{1},\ldots,h_{D}\}$ for some
finite $D$ (unspecified). Then~$N_{t}$ is the scale load of $h_{t}$ and the
population mean is expressed in terms of the scale loads as $\bar{Y}
=\break
N^{ - 1}\sum_{t = 1}^{D} N_{t} h_{t}$. Reduced sample data under simple
random sampling without replacement is represented by the sample scale
loads $n_{t},t = 1,\ldots,D$, and the resulting likelihood function is the
hyper-geometric likelihood $L(N_{1},\ldots,N_{D})$. If the sampling fraction
is negligible, then the likelihood is simply the multinomial likelihood
which is the same as the empirical likelihood (EL) of Owen (\citeyear{Owe88}). In
the~ca\-se of stratified random sampling, the likelihood function is the
product of hyper-geometric likelihoods corresponding to the different
strata $h = 1,\ldots,L$.

Hartley and Rao focused primarily on design-ba\-sed inferences, but also
briefly studied Bayesian inference under simple random sampling using a
com\-pound-multinomial prior on the scale loads $N_{1},\ldots,\break N_{D}$. Hoadley
(\citeyear{Hoa69}) obtained the compound-multi\-nomial prior,
denoted $\operatorname{CMtn}(N_{d};\nu_{d},d = 1,\ldots,D)$, by first assuming that the
finite population $\{ y_{i},i \!=\! 1,\ldots,\break N\}$ is a random sample from an
infinite population with unknown probabilities $p_{d} = P(y_{i} =
h_{d}),d =\break 1,\ldots,D$, and then using a Dirichlet prior with parameters
$\nu_{d}$ ($ \!>\! 0$) on the probabilities $p_{d},d \!=\! 1,\ldots,D$. The posterior
distribution of $N_{d} - n_{d}$, $d = 1,\ldots,D$, given the data $n_{d}$, $d =
1,\ldots,D$, is the compound multinomial $\operatorname{CMtn}(N_{d} - n_{d};\nu_{d} +
n_{d},d = 1,\ldots,D)$. Using this posterior distribution, Hartley and Rao
(\citeyear{HARRao68}) obtained the posterior mean and the posterior variance of the
population mean $\bar{Y}$. Under a diffuse prior with the $v_{d}$ close to
zero, the results are identical to those of Ericson (\citeyear{Eri69}). However,
there are fundamental differences in the two approaches in the sense
that under exchangeability the conditional distribution of the sample
scale loads $n_{d}$, given the population scale loads $N_{d}$, is equal to
the hyper-geometric likelihood of Hartley--Rao for any sampling design,
whereas in the Hartley--Rao approach this conditional distribution and
the resulting posterior of the $N_{d}$ are derived under simple random
sampling, and hence depend on the sampling design. Rao and Ghangurde
(\citeyear{RaoGha72}) studied Bayesian optimal sample allocation, by minimizing the
expected posterior variance of the mean, for stratified simple random
sampling and some other cases including two-phase sampling to handle the
nonresponse problem. Attention was given to data-based priors obtained
by combining diffuse priors with likelihoods based on pilot samples.

Aitkin (\citeyear{AIT08}) used the scale-load framework and obtained Bayesian
intervals on the population mean under simple random sampling, by using
a com\-pound-multinomial prior with $v_{d} \!=\! 0$ on the observed scale loads
$n_{d} > 0$ and then simulating a large number of samples from the
resulting posterior distribution. This approach is similar to the
simulation approach used by Meeden and Vardeman (\citeyear{MeeVar91}), but the
posterior intervals depend on the design via the likelihood function. As
in the case of Meeden and Vardman, the simulation method can be applied
to other parameters of interest. Also, the simulation method readily
extends to stratified simple random sampling.

The scale-load approach is promising but somewhat limited in
applicability, in the sense that the scale-load likelihoods cannot be
obtained easily for complex sampling designs.
To handle complex sampling designs, Rao and Wu (\citeyear{RAOWU10}) used a pseudo-EL
approach, proposed by Wu and Rao (\citeyear{WuRao06}), to obtain ``calibrated''
pseudo-Bayesian intervals in the sense that the intervals have
asymptotically correct designbased coverage probabilities. The pseudo-EL
approach uses the survey weights and the design effect (via the
effective sample size $n^{*}$) in defining\vspace*{2pt} the profile pseudo-EL function
for the mean $\theta = \bar{Y}$. Let $\tilde{d}_{i}(s) = d_{i}/\sum_{j
\in s} d_{j}$ be the normalized weights and $n^{*} = n/(\mathrm{deff})$, where
deff is the ratio of the estimated variance of the weighted mean
$\sum_{i \in s} \tilde{d}_{i} (s)y_{i}$ to
the estimate of the variance
under simple random sampling. Then\vadjust{\goodbreak} the profile pseudo empirical
log-likelihood function for $\theta$ is given by
\begin{equation}\label{e4}
l_{\mathrm{PEL}}(\theta) = n^{*}\sum_{i \in s} \tilde{d}_{i} (s)\log\{
\hat{p}_{i}(\theta)\},
\end{equation}
where the $\hat{p}_{i}(\theta)$ maximize $\sum_{i \in s} \tilde{d}_{i}
(s)\log p_{i}$ subject to $p_{i} > 0,\sum_{i \in s} p_{i} = 1$
and $\sum_{i \in s} p_{i} y_{i} = \theta$. We refer the reader to Rao and
Wu (\citeyear{autokey83}) for an overview of EL methods used for inference from survey
data.

By combining the profile pseudo-EL function for the population mean from
(\ref{e4}) with a flat prior on the mean, one can get pseudo-Bayesian intervals
that have asymptotically correct design-based coverage probabilities.
Also, it may be easier to specify informative priors on the mean if
historical information on the mean is available. The proposed approach
can incorporate known auxiliary population information in the
construction of pseudo-Bayesian intervals using the basic design weights
or using weights already calibrated by the known auxiliary information.
The latter is more appealing because, in practice, data files report the
calibrated weights. One limitation of the Rao--Wu method for complex
designs is that the pseudo-EL depends on the design effects which may
not be readily available. Lazar (\citeyear{Laz03}) proposed the Bayesian profile EL
approach for the case of independent and identically distributed (i.i.d.)
observations.

It should be noted that even in the i.i.d. case a~``matching'' prior on the
mean that provides higher order coverage accuracy for the intervals does
not exist when using the nonparametric Bayesian profile-EL (Fang and
Mukerjee, \citeyear{FanMuk06}). Therefore, the main advantage of the approach is to get
``exact'' pseudo-Bayesian intervals that are also calibrated in the
sense of first order coverage accuracy in the design-based framework.

\section{Small Area Estimation: HB Approach}\label{5}

Methods for small area (or domain) estimation have received much
attention in recent years due to growing demand for reliable small area
statistics. Traditional area-specific direct estimation methods (either
design-based or model based or Bayesian) are not suitable in the small
area context because of small (or even zero) area-specific sample sizes.
As a result, it is necessary to use indirect estimation methods that
borrow information across related areas through linking models based on
survey data and auxiliary information, such as recent census data and
current administrative records. Advocates of design-based methods indeed
acknowledge the need for models in small area estimation. For example,
\citeauthor{HANMADTEP83} (\citeyear{HANMADTEP83}) remark, ``If the assumed model accurately
represents the state of nature, useful inferences can be based on quite
small samples at least for certain models.''

Linking models based on linear mixed models and generalized linear mixed
models with random small area effects are currently used extensively, in
conjunction with empirical best linear unbiased prediction (EBLUP),
parametric empirical Bayes (EB) and hierarchical Bayes (HB) methods for
estimation of small area means and other small area parameters of
interest. A detailed treatment of small area estimation methods is given
in Rao (\citeyear{Rao03}). We focus here on HB methods to highlight the significant
impact of Bayesian methods on small area estimation.

In the HB approach, model parameters are treated as random variables and
assigned a prior distribution. Typically, noninformative priors are
used, but one must make sure that the resulting posteriors are proper
because some priors on the variance parameters can lead to improper
posteriors (see Rao, \citeyear{Rao03}, Section 10.2.4, for a discussion on the
choice of priors). The posterior distribution of a small area parameter
of interest is then obtained from the prior and the likelihood function
generated from the data and the assumed model. Typically, closed-form
expressions for desired posterior distributions do not exist, but
powerful MCMC methods are now available for simulating samples from the
desired posterior distribution and then computing the desired posterior
summaries. Rao (\citeyear{Rao03}, Chapter~10) gives a detailed account of the HB
methods in the small area context; see also the review paper by Datta
(\citeyear{DAT09}), Section~3.

A significant advantage of the HB approach is that it is straightforward
and the inferences are ``exact,'' unlike in the EB approach. Moreover,
it can handle complex small area models using MCMC methods. Availability
of powerful MCMC methods and software, such as WinBUGS, also makes HB
attractive to the user. Extensive HB model diagnostic tools are also
available, but some of the default HB model-checking measures that are
widely used may not be necessarily good for detecting model deviations.
For example, the commonly used posterior predictive $p$-value (PPP) for
checking goodness of fit may not be powerful enough to detect
nonnormality of random effects (Sinharay and Stern,  \citeyear{SinSte03}) because this
measure makes ``double use'' of data in the sense of first generating
values from the predictive posterior distribution and then calculating
the $p$-value. Bayarri and Castellanos (\citeyear{BayCas07}) say, ``Double use of the
data can result in an extreme conservatism of the resulting $p$-values.''\vadjust{\goodbreak}
Alternative measures, such as the partial PPP and the conditional PPP
(Bayarri and Berger, \citeyear{BayBer00}), attempt to avoid double use of data, but
those measures are more difficult to implement than the PPP, especially
for the small area models. Browne and Draper (\citeyear{BRODRA}) suggested the use of
prior-free, frequentist methods in the model exploration phase and then
the HB for inference based on the selected models using possibly diffuse
priors on the model parameters. However, many Bayesians may not agree
with this suggestion because of the orientation of frequentist tests of
goodness of fit to rejecting null hypotheses, as noted by a referee.

To illustrate the HB approach for small area estimation, we focus on a
basic area-level model with two components, a sampling model and a
linking model, requiring only area-specific (direct) designbased
estimators $\bar{y}_{iw}$ of small area means $\bar{Y}_{i}$ and
associated area-level covariates $z_{i}$ ($i = 1,\ldots,m$). The linking model
is of the form $g(\bar{Y}_{i}) = z'_{i}\beta + v_{i}$, where the random
effects $v_{i}\sim_{\mathrm{i.i.d.}}N(0,A)$ and $g(\cdot)$ is a~specified link function.
The sampling model assumes that $g(\bar{y}_{iw}) = g(\bar{Y}_{i}) +
\tilde{e}_{i}$, where the sampling errors $\tilde{e}_{i}|\bar{Y}_{i}$
are assumed to be independent $N(0,D_{i})$ with known sampling
variances $D_{i}$. The assumptions of zero mean sampling errors and known
sampling variances may be both quite restrictive in practice. The first
difficulty may be circumvented by using the sampling model $\bar{y}_{iw}
= \bar{Y}_{i} + e_{i}$, where the sampling errors $e_{i}$ are assumed to
be independent normal with zero means, which simply says that the direct
estimators are design unbiased or nearly design unbiased, as in the case
of a GREG estimator, for large overall sample size. The second
assumption of known sampling variances is more problematic and the usual
practice to get around this problem is to model the estimated sampling
variances (using generalized variance functions) and then treat the
resulting smoothed estimates as the true variances $D_{i}$. Bell (\citeyear{BEL08})
studied the sensitivity of small area inferences to errors in the
specification of the true variances. The original model, called the
Fay--Herriot (FH) model, is a~matched model in the sense that the
sampling model matches the linking model and the combined model is
simply a special case of a linear mixed model. On the other hand, the
alternative sampling model is not necessarily matched to the linking
mo\-tirdel and in this case the two models are ``mismatched.'' For
simplicity, we focus on the matched case, but the HB approach readily
extends to the more complex case of mismatched models and also to models
that allow the sampling variance to depend on the area mean $\bar{Y}_{i}$
(You and Rao, \citeyear{YouRao02}).

Attractive features of area level models are that the sampling design is
taken into account through the direct estimators $\bar{y}_{iw}$ and that
the direct estimators and the associated area level covariates are more
readily available to the users than the corresponding unit level sample
data. For example, the U.S. Small Area Income and Poverty Estimation
(SAIPE) Program used the FH model to estimate county level poverty
counts of school-age children by employing direct estimates for sampled
counties from the Current Population Survey and associated county level
auxiliary information from tax records, food stamps programs and other
administrative sources (see Rao, \citeyear{Rao03}, Chapter 7, for \mbox{details}).
Bayesians have used the area level models extensively through the HB
approach, in spite of the limitations mentioned above, because of their
practical advantages (Rao, \citeyear{Rao03}, Chapter 10).

In the HB approach, a flat prior on the model parameters $\beta$ and
$A$ is often specified as $f(\beta,A) \propto f(A)$ and $f(A) \propto 1$,
and the resulting posterior summaries (means, variances and credible
intervals) for the means $\bar{Y}_{i}$ are obtained. Bell (\citeyear{BEL}) studied
ma\-tched models in the context of estimating the proportion of school-age
children in poverty at the state level in the US, using the survey
proportions $\bar{y}_{iw} = p_{iw}$ based on the Current Population
Survey (CPS) data for 1989--1993 and area level covariates $z_{i}$
related to $\bar{Y}_{i} = P_{i}$. Bell found that the maximum likelihood
(ML) and restricted ML (REML) estimates of $A$ turned out to be zero for
the first four years (1989--1992) and the resulting EB estimates of state
poverty rates attached zero weight to the direct estimate $p_{iw}$
regardless of the CPS state sample sizes~$n_{i}$ (number of households).
This problem with EB ba\-sed on ML or REML can be circumvented by using
the HB approach. Bell used the above flat prior and obtained the
posterior mean which always attached nonzero weight to the direct
estimate. Further, the posterior variance is well behaved (smallest for
California with the largest $n_{i})$, unlike the estimated mean squared
prediction error (MSPE) of the EB estimator. It is possible, however, to
develop EB methods that always lead to nonzero estimates of $A$. Morris
(\citeyear{JiaLah06N2}) proposed to multiply the residual likelihood function of $A$ by
the factor $A$ and maximize this adjusted likelihood function. The
resulting estimator of $A$ is always positive and gets around the
difficulty with REML. Li and Lahiri (\citeyear{LiLah10}) used an adjusted profile
likelihood function which also leads to positive estimates of $A$. They
also established asymptotic consistency of the estimator and obtained a
nearly unbiased estimator of the mean squared prediction error (MSPE) of
the associated EB estimator of $\bar{Y}_{i}$.\vadjust{\goodbreak}

\citeauthor{DatRaoSmi05} (\citeyear{DatRaoSmi05}) studied frequentist properties of HB by deriving a
moment-matching prior on $A$, in the sense that the resulting posterior
variance is nearly unbiased for the MSPE of the HB estimator of the
small area mean. The moment-matching prior is given by
\begin{equation}\label{e5}
f(A) \propto(A + D_{i})^{2}\sum_{l = 1}^{m} (A + D_{l} )^{ - 2}.
\end{equation}
This prior depends collectively on the sampling variances $D_{l}$ for all
the areas as well as on the area-specific sampling variance$D_{i}$. Note
that the matching prior is designed for inference on area $i$ and,
hence, its dependence on $D_{i}$ should not be problematic. The matching
prior (\ref{e5}) reduces to the flat prior $f(A) \propto 1$ in the special case
of equal sampling variances $D_{i} = D$. However, in the application
considered by Bell (\citeyear{BEL}), $\max D_{i}/\min D_{i}$ is as large as 20.
Ganesh and Lahiri (\citeyear{GanLah08}) derived a single matching prior such that a
weighted posterior variance over the areas tracks the corresponding
weighted MSPE for specified weights. By letting the weights be one for
area $i$ and zero for the remaining areas, the resulting prior is
identical to~(\ref{e5}). Datta (\citeyear{DAT}) has shown that the previous
moment-matching priors also ensure matching property for interval
estimation in the sense that the coverage probability of the credible
interval tracks the corresponding coverage probability of the normal
interval based on the EB estimator and its estimated MSPE. Further work
on matching priors in the context of small area estimation would be
useful.

Mismatched models are often more realistic for practical applications,
as they allow flexibility in formulating the linking model. A recent
application of HB under mismatched models is to the estimation of adult
literacy levels for all states and counties in the US, using data from
the National Assessment of Adult Literacy and literacy-related auxiliary
data (Mohadjer et al., \citeyear{MOHetal}). Bizier et al. (\citeyear{BIZetal}) used mismatched
models and the HB approach to produce estimates of disability rates for
health regions and selected municipalities in Canada.

A variety of applications of HB under complex modeling have been
reported in the literature (see Rao, \citeyear{Rao03}, Chapter 10, for work prior to
2003). Nandram and Choi (\citeyear{NANCHO05}) and \citeauthor{NANCOXCHO05} (\citeyear{NANCOXCHO05}) studied
extensions of HB to handle nonignorable nonresponse and applied the
methods to data from the National Health and Nutrition Examination
Survey (NHANES III) to produce small area estimates. Raghunathan et al.
(\citeyear{Ragetal07}) applied the HB approach to combine data from two independent
surveys\vadjust{\goodbreak} [Behavioral Risk Factor Surveillance System (BRFSS) and the
National Health Interview Survey (NHIS)] for the years 1997--2000 to
produce yearly prevalence estimates at the county level for six
outcomes. BRFSS is a large telephone survey covering almost all US
counties, but the nonresponse rates are high and also nontelephone
households are not covered. On the other hand, NHIS is a smaller
personal interview survey with lower nonresponse rates and covers
nontelephone households. In this application, direct survey weighted
county estimates of proportions from the two surveys were transformed
using the inverse sine transformation and the sampling variances were
taken as $(4\tilde{n}_{d})^{ - 1}$, where $\tilde{n}_{d}$ denotes the
effective sample size for a particular domain $d$ (calculated as the
actual domain sample size~$n_{d}$ divided by the estimated design effect
which is the ratio of the estimated variance under the given design to
the binomial estimated variance). The resulting sampling model was then
combined with a~suitable linking model to obtain county estimates of the
prevalence rates, using diffuse proper priors on the model parameters
and MCMC. This application attempts to account for possible noncoverage
bias and obtain efficient county estimates by combining data from two
independent surveys. It may be noted that the model used here is an
extension of the basic FH area level model and the application
demonstrates how design-based and Bayesian approaches can be fruitfully
integrated in small area estimation.

HB methods studied in the literature have been largely parametric, based
on specified distributions for the data. However, Meeden (\citeyear{MEE03}) extended
his noninformative Bayesian approach, based on the Po\-lya posterior (PP),
to small area estimation in some simple cases. Extension of this
approach to handle complex models is not likely to be easy in practice.

\section{Concluding Remarks}

I have provided an appraisal of the role of Bayesian and frequentist
methods in sample surveys. My opinion is that for domains
(subpopulations) with sufficiently large samples, a traditional
design-based\vadjust{\eject} frequentist approach that makes effective use of auxiliary
information, through calibration or assistance from working models, will
remain as the preferred approach in the large-scale production of
official sta\-tistics from complex surveys. Nonsampling errors can be
handled using a combined design and model approach with minimal use of
models. But the designbased approach, using survey weights, is not a~%
pa\-nacea even for large samples and yet ``many people ask too much of the
weights'' (Lohr, \citeyear{Loh07}), prompting statements like, ``Survey weighting is
a mess'' (Gelman, \citeyear{Gel07}). As Lohr (\citeyear{Loh07}) noted, survey weighting is not a
mess as long as the weighting is not stretched to a limit as in the case
of a very large number of post-stratified cells leading to very small or
even zero cell sample sizes, thus making weighting at the cell level
unstable or even not feasible (Gelman, \citeyear{Gel07}). Alternative weighting
methods can be used in those situations to get around this problem. For
example, by calibrating to the marginal counts of the
post-stratification variables instead of the cell counts leads to a
calibration estimator with stable weights which should perform well for
estimating population totals or means. Also, the resulting weights do
not depend on the response values, thus ensuring internal consistency,
unlike the hierarchical regression method proposed by Gelman (\citeyear{Gel07})
based on models involving random effects.

Recent work on nonparametric Bayesian methods that can be used for both
Bayesian and design-based inferences looks promising, at least for some
specialized surveys. For small area estimation, the hierarchical Bayes
(HB) approach offers a lot of promise because of its ability to handle
complex small area models and provide ``exact'' inferences. However, the
choice of noninformative priors that can provide frequentist validity is
not likely to be easy in practice when complex modeling is involved.
Also, caution needs to be exercised in the routine use of popular HB
model-checking methods.

\section*{Acknowledgments}

This research was supported by a grant from the Natural Sciences and
Engineering Research Council of Canada. My thanks are due to an
associate editor and two referees for constructive comments and
suggestions.


\end{document}